\documentclass[11pt]{article}
\usepackage{aas2pp4}
\slugcomment{Accepted for publication in the Astrophysical Journal}

\newcommand{\chisq}{$\chi^2$}
\newcommand{\eg}{e.g.}
\newcommand{\etal}{et~al.}
\newcommand{\gammaray}{$\gamma$-ray}
\newcommand{\gammarays}{$\gamma$-rays}
\newcommand{\perareasec}{~cm$^{-2}$~s$^{-1}$}
\newcommand{\perareasecmev}{~cm$^{-2}$~s$^{-1}$~MeV$^{-1}$}

\newcommand{\eq}[1]{equation~(\ref{#1})}
\newcommand{\fig}[1]{Figure~\ref{#1}}
\newcommand{\tbl}[1]{Table~\ref{#1}}

\begin{document}
\title{Phase-Resolved Studies of the High-Energy Gamma-Ray Emission from
the Crab, Geminga, and Vela Pulsars}
\author{J.~M.~Fierro, P.~F.~Michelson, P.~L.~Nolan}
\affil{W. W. Hansen Experimental Physics Laboratory,
Stanford University, Stanford, CA 94305-4085}

\author{D.~J.~Thompson}
\affil{NASA/Goddard Space Flight Center, Code 661, Greenbelt, MD 20771}

\begin{abstract}
Using the first three and a half years of observations from the Energetic
Gamma Ray Telescope (EGRET) on board the Compton Gamma Ray Observatory
(CGRO), phase-resolved analyses are performed on the emission from the three
brightest high-energy \gammaray\ pulsars, Crab, Geminga, and Vela.  For each
pulsar, it is found that there is detectable high-energy \gammaray\ emission
above the galactic diffuse background throughout much of the pulsar rotation cycle.
A hardness ratio is introduced to characterize the evolution of the spectral
index as a function of pulsar phase.  While the hardest emission from the
Crab and Vela pulsars comes from the bridge region between the two \gammaray\
peaks, the hardest emission from Geminga corresponds to the second \gammaray\
peak.  For all three pulsars, phase-resolved spectra of the pulse profile
components reveal that although there is a large variation in the spectral
index over the pulsar phase interval, the high-energy spectral turnover, if
any, occurs at roughly the same energy in each component.  The high-energy
\gammaray\ emission from the Crab complex appears to include an unpulsed
ultra-soft component of spectral index $\sim -4.3$ which dominates the total
emission below 100 MeV\@.  This component is consistent with the expected
emission from the tail end of the Crab nebula synchrotron emission.
\end{abstract}


\section{Introduction}
High-energy
\gammaray\ pulsars have proven to have complex, often extended, pulse
profiles.  By extracting only those photons which arrive during a specified
phase interval, the emission characteristics of an individual pulse pulse
profile component can be determined, providing there are sufficient photon
statistics.  Fortunately, the three brightest compact sources in the
\gammaray\ sky are the high-energy \gammaray\ pulsars Crab (PSR B0531+21),
Geminga (PSR J0633+1746), and Vela (PSR B0833$-$45).  The spectral properties
of the pulse profile components from these pulsars have been investigated
previously
using data from both COS B (\cite{Clear87}; \cite{Grenier88}) and EGRET
(\cite{Nolan93}; \cite{Kanbach94}; \cite{Mayer94}).

The earlier analyses for Crab and Vela compared the angular distribution
of the data with the instrument point-spread distribution to define the
source flux; no use was made of a detailed background model.  Thus the
non-source background level was not well determined.  In the Geminga
analysis by Mayer-Hasselwander et al. (1994), the non-source background
in the pulse profile was determined by spatial (cross correlation)
analysis of the counts in the off-pulse phase interval, simultaneously
accounting for the model-predicted diffuse emission.
In the present analysis, the increase in photon statistics available
through the first 3.5 years of EGRET observations allows 20 phase
intervals to be analyzed individually for all three pulsars by a 
spatial analysis procedure based on the maximum-likelihood concept
(\cite{Mattox96}), where the non-source background has been accounted
for on the basis of an improved model of the diffuse emission
(\cite{Hunter97}).  Some differences from earlier results may be caused
by differences in the analysis techniques.  Within this analysis each
pulsar is analyzed by the same procedure, making their results directly
comparable.

An improved knowledge of the emission characteristics as a function of pulsar
rotation phase will lead to a better understanding of the emission processes
and pulsar geometries responsible for modulated high-energy \gammarays.  In
particular, the phase-resolved behavior may be used to discriminate between
the two most popular classes of models for pulsed high-energy
$\gamma$-radiation: the polar cap models (\eg\
\cite{Daugherty94,Daugherty96}; \cite{Sturner95a}; \cite{Sturner95b}), which
propose that \gammarays\ result from energetic charged particles accelerated
along the magnetic field lines just above the polar cap surfaces, and the
outer gap models (\eg\ \cite{Cheng86a,Cheng86b}; \cite{Romani95};
\cite{Yadigaroglu95}), in which it is postulated that \gammarays\ are
produced by charged particles accelerated in vacuum gaps formed in the outer
regions of the pulsar magnetosphere.  Polar cap $\gamma$-radiation should be
narrowly beamed in the same direction as the radio emission, with a possible
high-energy cutoff in the photon spectrum due to attenuation by the strong
magnetic field near the polar cap surface.  The outer gap \gammaray\ beam is
expected to be broad and complex, consisting of emission generated along the
last closed field lines.

This paper examines the phase-resolved emission characteristics of the three
brightest high-energy \gammaray\ pulsars Crab, Vela, and Geminga.  A limited
phase-resolved analysis of the high-energy \gammaray\ emission from
PSR B1706$-$44 is presented by Thompson \etal\ (1996).  The remaining two
pulsars detected by EGRET, PSR B1055$-$52 and PSR B1951+32, do not yet have sufficient photon counting statistics to perform a meaningful phase-resolved
study of their emission (\cite{Fierro95}).

\section{Observations}
EGRET is sensitive to \gammarays\ in the energy range from $\sim$ 30 MeV to
$\sim$ 30 GeV\@.  The EGRET instrument and its calibration have been
described extensively in Hughes \etal\ (1980), Kanbach \etal\ (1988, 1989),
Nolan \etal\ (1992), and Thompson \etal\ (1993).  EGRET records each
\gammaray\ as an electron-positron pair production event.  This event is
processed automatically to determine the optimal estimate of arrival
direction and energy of the photon (\cite{Bertsch89}).  The arrival time of
each photon is recorded in Universal Coordinated Time (UTC) with an absolute
timing accuracy of better than 100 $\mu$s.  Because of the very low flux
level of high-energy \gammarays, observing periods typically last about two
weeks.

This paper considers the combined observations made during the first three 
cycles of the CGRO mission, which lasted from 1991 April to 1994 October.  
As shown by Ramanamurthy et al. (1995b), Mayer-Hasselwander et al.
(1994), and McLaughlin et al. (1996), 
these pulsars show no evidence of strong variability.  The present data 
include the results of  the in-flight calibration analysis of EGRET 
(\cite{Esposito97}).  This analysis added a small energy-dependent 
correction after the data set used by Ramanamurthy et al. (1995b) was complete. 

 The photon
arrival times were transformed to solar system barycentric time $T$, and
the corresponding rotational phases $\phi$ for each pulsar were determined by
taking the fractional part of the Taylor expansion
\begin{eqnarray}
\phi(T) &= \phi(T_0) + f_0 \, (T-T_0) + \frac{1}{2} \, f_1 \, (T-T_0)^2 
\nonumber \\
&\quad + \frac{1}{6} \, f_2 \, (T-T_0)^3 \; ,
\label{phase}
\end{eqnarray} 
where $f_0$, $f_1$, and $f_2$ are the pulsar spin frequency and first two
time derivatives measured at the reference epoch $T_0$.  The ephemerides used
for the temporal analysis of the Crab, Geminga, and Vela pulsars are listed
in Tables~\ref{crab_ephem}, \ref{geminga_ephem}, and \ref{vela_ephem},
respectively.  The epoch is given in units of Modified Julian Days (MJD = JD $-$ 2400000.5).
The Crab and Vela timing solutions are made available via the regularly
updated Princeton Pulsar Timing Database and its associated corrections for dispersion measure drifts(\cite{Arzoumanian92}).  The value of
$\Delta T_\oplus$ listed in the last column of Tables~\ref{crab_ephem} and
\ref{vela_ephem} is the amount of time after 0$^{\rm h}$ UTC on the epoch day
that a main radio pulse would arrive at the center of the Earth if there were
no dispersion delays.  This quantity is used to determine $\phi(T_0)$ in
\eq{phase}.  
Since Geminga is not a radio pulsar, its timing solution is derived using 
a ``$Z_n^2$ timing method'' based on the first four years of EGRET observations
(\cite{Mattox94,Mattox95}).  The frequency second derivative $f_2$ of
Geminga is too small to be measured and is set equal to zero.

\begin{deluxetable}{lccccc}
\tablecaption{Radio Timing Parameters of the Crab Pulsar\label{crab_ephem}}
\tablewidth{0pt}
\tablehead{
\colhead{} & \colhead{$T_0$} & \colhead{$f_0$} & \colhead{$f_1$} & 
\colhead{$f_2$} & \colhead{$\Delta T_\oplus$} \\
\colhead{Valid Dates} & \colhead{(MJD)} & \colhead{(s$^{-1}$)} &
\colhead{($10^{-10}$~s$^{-2}$)} & \colhead{($10^{-20}$~s$^{-3}$)} &
\colhead{(ms)}
}
\startdata
1991 Mar 16--Jun  5      & 48371 & 29.9492515379593 & -3.77657 & 0.818 &  8.3 \nl
1991 May 30--Aug 25      & 48449 & 29.9467067038240 & -3.77575 & 1.06  & 23.8 \nl
1992 Jun 28--Oct 18      & 48857 & 29.9334039601096 & -3.77156 & 1.16  &  1.5 \nl
1992 Dec 27--1993 Apr 11 & 49035 & 29.9276050289288 & -3.76972 & 1.05  & 25.8 \nl
1993 Apr  6--Jul 5       & 49128 & 29.9245763662884 & -3.76877 & 1.01  & 15.5 \nl
1993 Nov  5--1994 Jan 27 & 49337 & 29.9177729429376 & -3.76650 & 1.32  & 23.0 \nl
1993 Dec 19--1994 Feb 28 & 49375 & 29.9165363960966 & -3.76612 & 0.921 &  4.9 \nl
1994 Jun 23--Nov 15      & 49598 & 29.9092823758179 & -3.76380 & 1.30  & 26.3 \nl
\enddata
\end{deluxetable}

\begin{deluxetable}{lcccc}
\tablecaption{Timing Parameters for the Geminga Pulsar\label{geminga_ephem}}
\tablewidth{0pt}
\tablehead{
\colhead{} & \colhead{$T_0$} & \colhead{$f_0$} & \colhead{$f_1$} &
\colhead{$f_2$} \\
\colhead{Valid Dates} & \colhead{(MJD)} & \colhead{(s$^{-1}$)} & 
\colhead{(s$^{-2}$)} & \colhead{(s$^{-3}$)}
}
\startdata
1989 Jul 8--1996 Apr 24 & 48750 & 4.2176690940300 & 
$-1.95206 \times 10^{-13}$ & $< 6 \times 10^{-24}$ \nl
\enddata
\end{deluxetable}

\begin{deluxetable}{lccccc}
\tablecaption{Radio Timing Parameters of the Vela Pulsar\label{vela_ephem}}
\tablewidth{0pt}
\tablehead{
\colhead{} & \colhead{$T_0$} & \colhead{$f_0$} & \colhead{$f_1$} & 
\colhead{$f_2$} & \colhead{$\Delta T_\oplus$} \\
\colhead{Valid Dates} & \colhead{(MJD)} & \colhead{(s$^{-1}$)} & 
\colhead{($10^{-11}$~s$^{-2}$)} & \colhead{($10^{-21}$~s$^{-3}$)} &
\colhead{(ms)}
}
\startdata
1991 Feb 1--May 22       & 48343 & 11.1988875613749 & -1.55793 & 1.18  & 66.2 \nl
1991 Aug 12--Sep 29      & 48504 & 11.1987003711301 & -1.56791 & 4.17  & 22.3 \nl
1991 Oct 21--1992 Jan 16 & 48593 & 11.1985799537572 & -1.56584 & 2.25  & 93.3 \nl
1993 Apr 1--Aug 8        & 49142 & 11.1978384490139 & -1.56166 & 0.876 & 84.3 \nl
1993 Jul 1--Aug 27       & 49197 & 11.1977642483505 & -1.56105 & 3.88  & 63.5 \nl
1993 Sep 16--Nov 20      & 49278 & 11.1976550192558 & -1.56063 & 1.08  & 84.8 \nl
1994 Aug 28--Sep 23      & 49605 & 11.1972260134043 & -1.56636 & 0.626 & 69.7 \nl
\enddata
\end{deluxetable}

\section{Phase-Resolved Spatial Analysis}
Analysis of \gammaray\ point sources in the EGRET dataset is typically
performed using a maximum likelihood technique (\cite{Mattox96}), in which a
\gammaray\ excess above the predicted background emission is required to be
spatially distributed as the instrument point spread function before being
considered as point source emission.  In the EGRET energy
range, the \gammaray\ background is assumed to consist of an isotropic,
extragalactic component and Galactic diffuse emission due primarily to
cosmic-ray particles interacting with matter and fields in the Galaxy
(\cite{Bertsch93}; \cite{Hunter97}).
Although the background model des not provide an adequate global fit to
the all-sky data, it is an acceptable representation of the sky
brightness in limited regions if the strengths of the galactic and
extragalactic components are treated as adjustable parameters.

The Crab, Vela, and Geminga pulsars have sufficient \gammaray\ counting
statistics that it is possible to restrict the spatial analysis to photons
arriving during a specific phase interval, allowing the absolute emission
levels to be determined across the pulse profile. 

The likelihood analysis is subject to small systematic uncertainties 
(\cite{Mattox96}), including inaccuracies in the assumed diffuse model, 
assumptions made about other sources in the field of view, and limitations 
of the EGRET calibration data.  Thompson et al. (1995), for example, 
recommend a 10\% systematic uncertainty be associated with any absolute number, 
and McLaughlin et al. (1996) find that a 6\% systematic uncertainty is 
sufficient to justify the assumption that the pulsars are not varying.  
For comparisons between different phases of a given pulsar, these systematic 
uncertainties can be neglected, because all phases are analyzed under the 
same conditions.

For each of the three bright \gammaray\ pulsars, the pulse profile is divided
into twenty equal-width phase bins ($\Delta \phi = 0.05$), and likelihood analysis is performed on the twenty resulting spatial photon counts maps.
There are not enough counts to perform a full spectral analysis for each of
the 20 phase bins.  Instead, a {\em hardness ratio} $R$ is defined as
\begin{equation}
R \equiv \frac{F(> {\rm 300~MeV})}{F({\rm 100-300~MeV})} \; ,
\label{hardness}
\end{equation}
where $F({\rm 100-300~MeV})$ is the likelihood flux measured between 100 and
300 MeV, and $F(> {\rm 300~MeV})$ is the likelihood flux measured above 300
MeV, both in units of photons\perareasec.  A source with a differential flux
behaving as a perfect power law above 100 MeV with a spectral index of
$-2.0$  will have a hardness ratio of $R \simeq 0.5$.

\subsection{The Crab Pulsar}
Analysis of the Crab region is complicated by the fact that a solar
flare bright enough to be detected by EGRET as a strong source occurred on 1991 June 11 (\cite{Kanbach93}), when
the Sun was in a direction only $\sim 4\fdg4$ from that of the Crab pulsar.
Rather than introduce possible errors by trying to account for the solar
flare emission, all photons arriving during the $\sim$8.5 hour flare period
are excluded from the Crab pulsar analysis.

\begin{figure}
\plotone{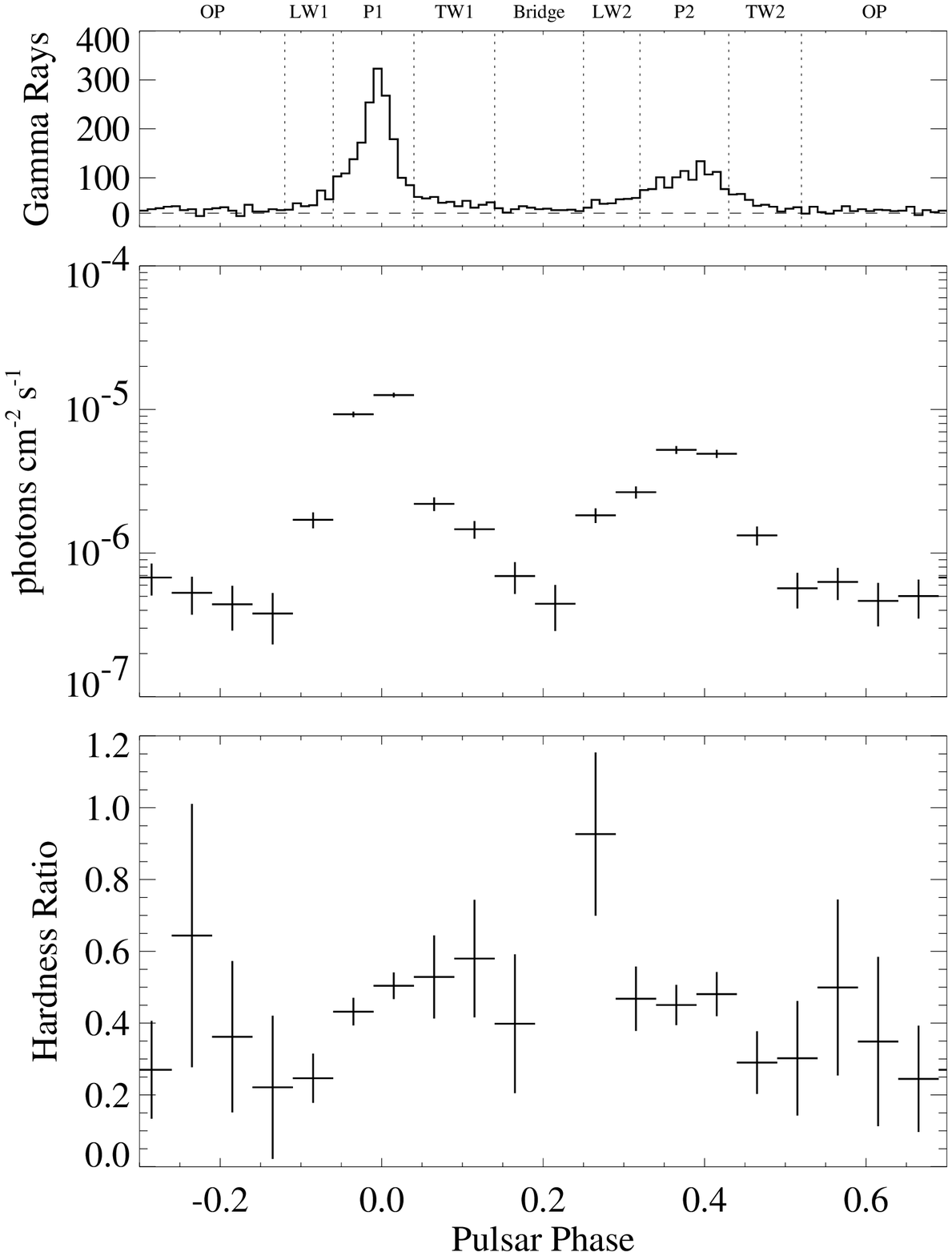}
\caption{Phase-resolved likelihood analysis of the Crab pulse profile.  
The top panel shows the \gammaray\ phase histogram above 100
MeV formed from the data accumulated during the first three cycles of EGRET
observations.  The vertical dashed lines indicate the definitions of the
pulse profile components listed in \tbl{crab_phasedef}.  The histogram
consists of 100 phase bins.  The pulsar phase is defined so that the main
radio peak occurs at zero.  The horizontal dashed line shows the level of non-source gamma radiation expected, based on spatial analysis of the off-pulse region.
The center panel shows instantaneous photon flux above the diffuse background level as derived from the spatial likelihood analysis of each phase bin for
energies greater than 100 MeV\@.
The lower panel plots the hardness ratios as defined by \eq{hardness}. 
\label{crab_hardness}}
\end{figure}

The results of the phase-resolved spatial likelihood analysis of the Crab
pulse profile are shown graphically in \fig{crab_hardness}.  The top panel
shows the traditional phase histogram above 100 MeV formed by epoch folding
all events from the first three cycles of EGRET observations arriving within
an energy-dependent cone of half-angle $\theta_{67} = 5\fdg85 \times
(E_\gamma/{\rm 100~MeV})^{-0.534}$ about the pulsar position, where
$E_\gamma$ is the measured photon energy.  This cone accepts $\sim$ 67\% of
the photons that EGRET detects from a point source (\cite{Thompson93}).  The
phase interval has been divided into the eight components defined in
\tbl{crab_phasedef}, indicated by dashed lines in the top panel of
\fig{crab_hardness}.  These boundary intervals differ from those of Clear
\etal\ (1987) and Nolan \etal\ (1993) because the increase in data has
allowed features to be resolved on a finer scale.  The horizontal dashed 
line represents the non-source background gamma radiation within this cone, 
determined as follows:  The off-pulse phase region was analyzed using the 
maximum likelihood method to calculate the number of photons associated with 
the Crab.  67\% of these should be in the light curve.  The non-source 
background is then the difference between the total number of photons and 
the number of source photons expected in this part of the light curve.  
This method is similar to the one used by Mayer-Hasselwander et al. (1994) 
for Geminga.  

\begin{deluxetable}{lcc}
\tablecaption{Crab Pulsar Phase Interval Definitions\label{crab_phasedef}}
\tablewidth{0pt}
\tablehead{
\colhead{Component} & \colhead{Name} & \colhead{Phase Interval}
}
\startdata
Leading Wing 1  & LW1    & 0.88 -- 0.94 \nl
Peak 1          & P1     & 0.94 -- 0.04 \nl
Trailing Wing 1 & TW1    & 0.04 -- 0.14 \nl
Bridge         	& Bridge & 0.14 -- 0.25 \nl
Leading Wing 2  & LW2    & 0.25 -- 0.32 \nl
Peak 2        	& P2     & 0.32 -- 0.43 \nl
Trailing Wing 2 & TW2    & 0.43 -- 0.52 \nl
Offpulse        & OP     & 0.52 -- 0.88 \nl
\enddata
\end{deluxetable}

The center panel of \fig{crab_hardness} shows the instantaneous photon flux
above 100 MeV measured for each of the twenty phase bins.  The total level of
emission from Crab is the average of the plotted values.  The error bars
represent the $1 \sigma$ statistical uncertainty levels.  The values are 
plotted on a
logarithmic scale because the photon flux varies by almost two orders of
magnitude across the pulse profile, with the single bin centered about the
first peak accounting for $\sim$ 35\% of the total Crab emission.  Unlike the
phase histogram shown in the top panel, the center plot shows only the
emission above the predicted \gammaray\ background level.  Significant
point-source emission is detected from the Crab region over almost the 
entire pulse profile. 
The low bin, centered on phase $-$0.15, has statistical significance of 
only 2.5$\sigma$, an excess of 27 photons on the expected non-source 
emission of 140.  All the other bins have significance greater than 
3.5$\sigma$.

Previous studies of the high-energy \gammaray\ emission from the Crab complex
(\cite{Clear87}; \cite{Nolan93}; \cite{Ramanamurthy95b}; \cite{deJager96})
have treated the emission in the offpulse
(OP) region of the pulse profile as an unpulsed component due entirely to the
Crab nebula.  This is in analogy with observational results at lower
energies, where the total Crab luminosity is dominated by the nebular
contribution.  Unlike detectors at lower energies, however, current
\gammaray\ telescopes do not have sufficient angular resolution to separate
the pulsar radiation from any possible nebular emission.  As will be shown in
the following sections, it is very likely that much of the emission in the
offpulse phase interval comes from the pulsar itself.  It is noteworthy that
the photon flux level measured for the phase bin at $\phi \approx 0.2$ is
comparable to the flux level measured over the offpulse phase interval.  It
had been assumed that the emission throughout the bridge region was well
above the offpulse emission level.  An unpulsed component present in the Crab
pulse profile can be no greater than flux measured in the weakest phase bin.
Thus, an unpulsed component has a maximum flux of $(3.8 \pm 1.5) \times
10^{-7}$ photons\perareasec, or $16 \pm 8\%$ of the total Crab emission above
100 MeV\@.

The lower panel of \fig{crab_hardness} plots the hardness ratios calculated
for the same set of phase bins.  There is no data point plotted for the bin
centered at phase $\phi \approx -0.15$ because the paucity of counts in that
bin prevented a detection in the 100--300 MeV energy range.  The most
striking aspect of this plot is the absence of a strong correlation to the
features in the pulse profile.  The hardness ratio increases smoothly as the
pulsar phase moves across the first pulse, leveling out in the bridge region
before reaching its maximum value in the leading wing of the second peak
($\phi \approx 0.25$).  A weak detection of this bin in the 100--300 MeV
energy range leads to the relatively large error bar.  As the phase increases
through the second peak, the hardness ratio decreases back to the soft level
of the offpulse emission.  It is interesting that the extreme hardness ratios
are in the bridge and offpulse phase intervals, not near the two peaks which
dominate the pulsar lightcurve.

\subsection{The Geminga Pulsar}
\begin{figure}
\plotone{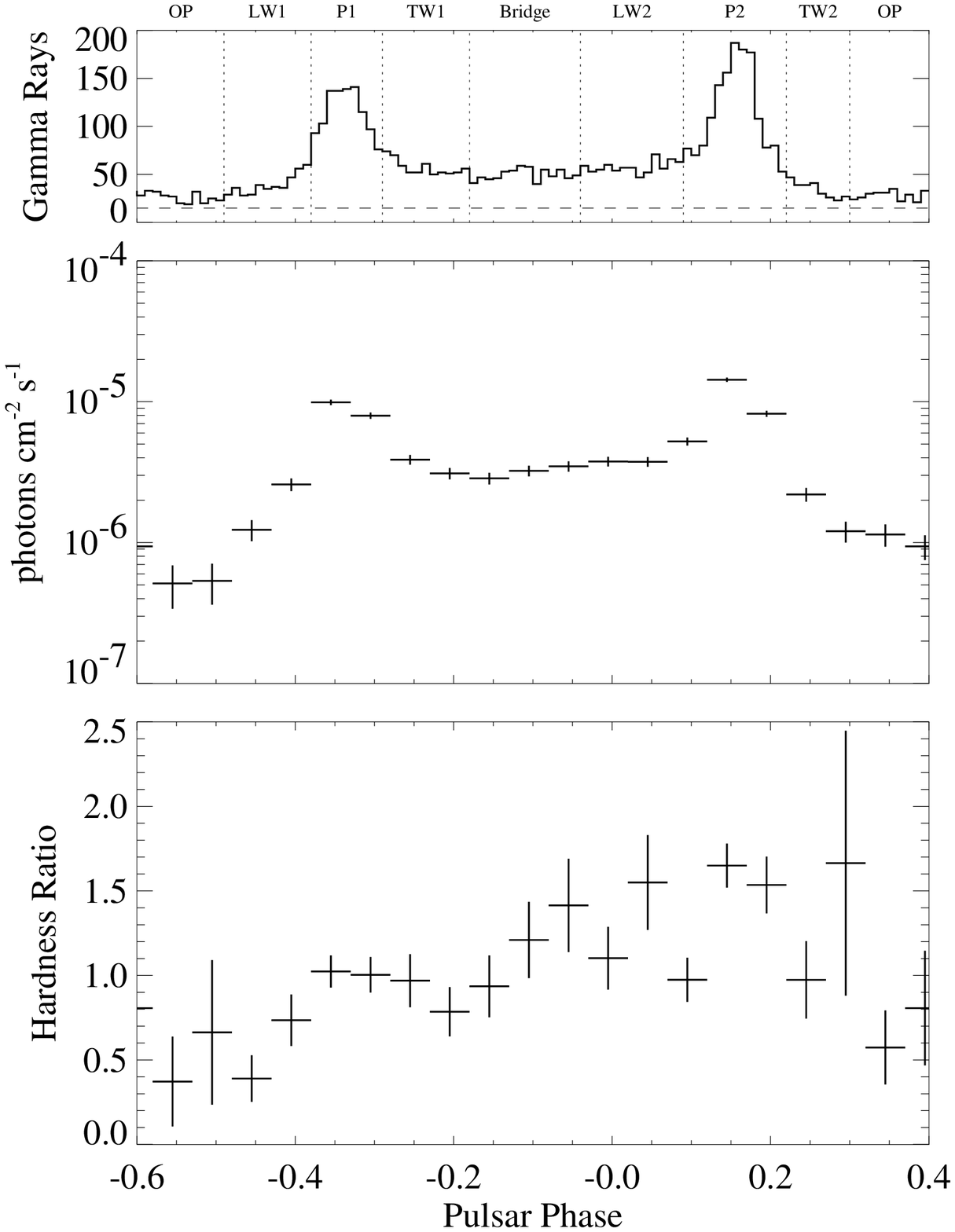}
\caption{Phase-resolved likelihood analysis of the Geminga pulse profile.  
The top panel shows the \gammaray\ phase histogram above 100 MeV, with the
vertical dashed lines indicating the definitions of the pulse profile
components listed in \tbl{geminga_phasedef}.  The histogram consists of 100
phase bins.  The pulsar phase is relative to the timing epoch. The horizontal dashed line shows the level of non-source gamma radiation expected, based on spatial analysis of the off-pulse region
The center panel shows instantaneous photon flux above the diffuse background
level for energies greater than 100 MeV\@, while the lower panel plots the hardness ratios. 
\label{geminga_hardness}}
\end{figure}

The phase histogram, phase-resolved instantaneous flux, and hardness ratio
distribution of Geminga are shown in \fig{geminga_hardness}.  The vertical
dashed lines in the top panel indicate the definitions of the pulse profile
components listed in \tbl{geminga_phasedef}, and the horizontal dashed line shows the non-source background emission, calculated in the same way as was done for the Crab.  Note that the dashed line is close to the observed points in the phase region leading the first pulse (cf. Fig. 2 of \cite{Mayer94}).
The plot of the instantaneous flux
level above 100 MeV shows that Geminga has an even higher average level of
offpulse emission than does the Crab.  However, unlike the Crab pulsar, Geminga
has no known nebular association, and it must be assumed that the high level
of emission seen throughout most of the pulse profile is coming directly 
from the pulsar, as noted previously by Mayer-Hasselwander et al. (1994).
The flux at phase $\phi \approx -0.55$ is $\sim 1/3$ the emission measured
from either of the adjacent phase bins, and, even considering the large
uncertainties, represents a significant drop in intensity.   This statistical 
significance of this bin is less than 2.5$\sigma$, so this point should 
conservatively be treated as an upper limit.  Any unpulsed
component to the total Geminga emission can not be stronger than the flux
level of $(5.1 \pm 1.7) \times 10^{-7}$ photons\perareasec\ measured in this
bin.  This corresponds to $14\% \pm 7\%$ of the total Geminga emission.  As
opposed to the Crab pulsar, the emission level over the entire bridge region
of the Geminga pulse profile is much higher than the offpulse emission
level.  Except for the one low bin, all of the phase regions show excesses at a statistical significance greater than 5$\sigma$.

\begin{deluxetable}{lcc}
\tablecaption{Geminga Pulsar Phase Interval Definitions\label{geminga_phasedef}}
\tablewidth{0pt}
\tablehead{
\colhead{Component} & \colhead{Name} & \colhead{Phase Interval}
}
\startdata
Leading Wing 1  & LW1    & 0.51 -- 0.62 \nl
Peak 1          & P1     & 0.62 -- 0.71 \nl
Trailing Wing 1 & TW1    & 0.71 -- 0.82 \nl
Bridge         	& Bridge & 0.82 -- 0.96 \nl
Leading Wing 2  & LW2    & 0.96 -- 0.09 \nl
Peak 2        	& P2     & 0.09 -- 0.22 \nl
Trailing Wing 2 & TW2    & 0.22 -- 0.30 \nl
Offpulse        & OP     & 0.30 -- 0.51 \nl
\enddata
\end{deluxetable}

The distribution of the Geminga hardness ratios, plotted in the lower panel
of \fig{geminga_hardness}, appears to be correlated to the basic features of
the \gammaray\ pulse profile.  There is perhaps a local maximum in the
hardness ratio at the location of the first peak, with even harder emission 
in the latter part of the bridge region and the second peak.
In general, the Geminga hardness ratios are
much higher than those measured for the Crab pulsar.

\subsection{The Vela Pulsar}
\begin{figure}
\plotone{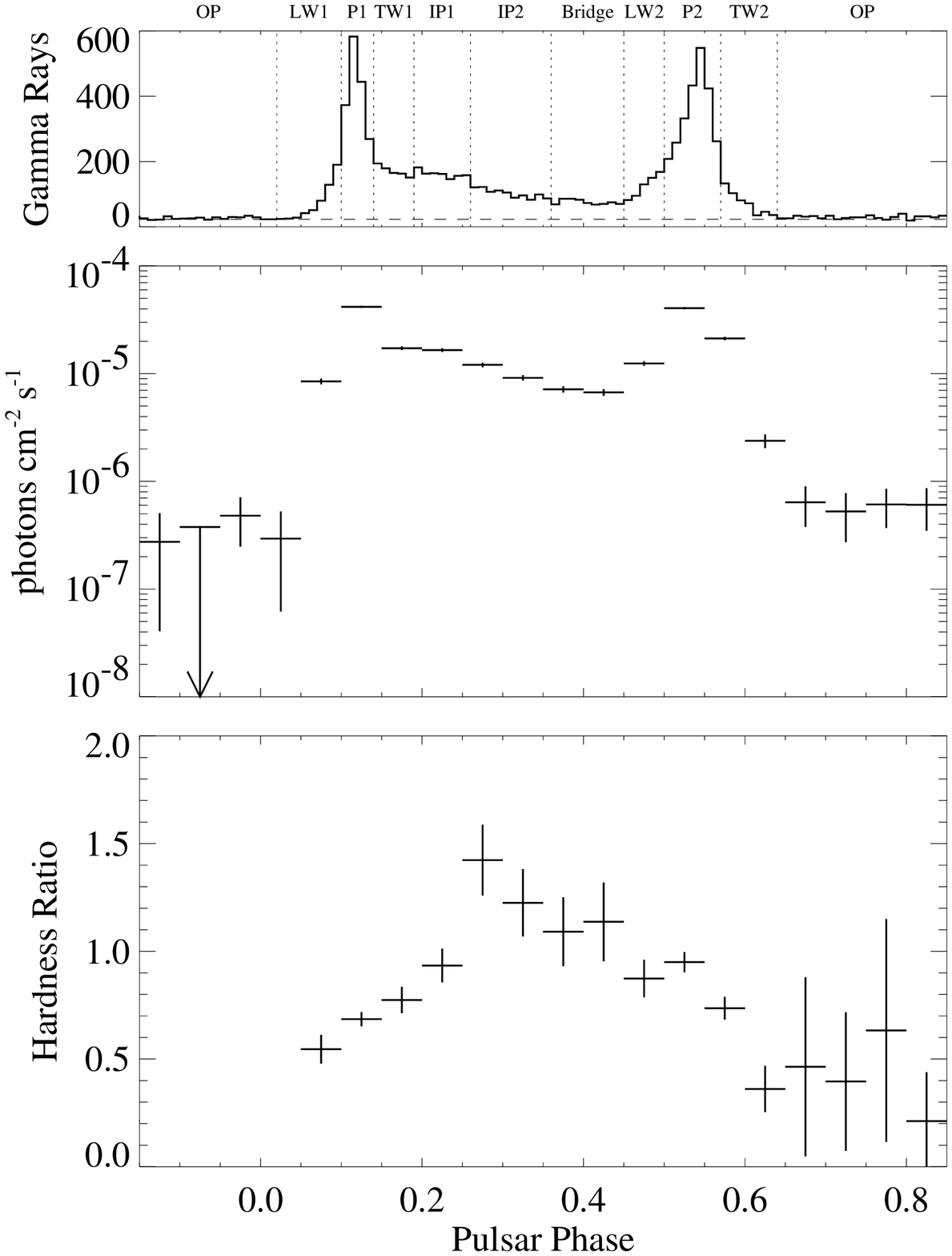}
\caption{Phase-resolved likelihood analysis of the Vela pulse profile.  
The top panel shows the \gammaray\ phase histogram above 100 MeV, with the
vertical dashed lines indicating the definitions of the pulse profile
components listed in \tbl{vela_phasedef}.  The histogram consists of 100
phase bins.  The pulsar phase is defined so that the main
radio peak occurs at zero. The horizontal dashed line shows the level of non-source gamma radiation expected, based on spatial analysis of the off-pulse region
The center panel shows instantaneous photon flux above the diffuse background
level for energies greater than 100 MeV\@, while the lower panel plots the hardness ratios. 
\label{vela_hardness}}
\end{figure}

\fig{vela_hardness} shows the phase histogram, phase-resolved flux, and
hardness ratios for the Vela pulsar, with the pulse profile components
defined in \tbl{vela_phasedef} indicated by vertical dashed lines in the top
panel. The horizontal dashed line shows the non-source background emission, calculated in the same way as was done for the Crab.  The increased statistics and complicated morphology
of the Vela pulse profile allows for the introduction of two additional
components not defined by Kanbach \etal\ (1994).  For lack of better names,
these have been denoted as ``interpulse 1'' (IP1) and ``interpulse 2'' (IP2),
and they correspond to the slight plateau after the first peak and its
trailing edge, respectively.  These are not to be confused with radio pulsar
astronomers' use of the term ``interpulse'' to denote the weaker pulse in a
two-pulse profile.

\begin{deluxetable}{lcc}
\tablecaption{Vela Pulsar Phase Interval Definitions\label{vela_phasedef}}
\tablewidth{0pt}
\tablehead{
\colhead{Component} & \colhead{Name} & \colhead{Phase Interval}
}
\startdata
Leading Wing 1  & LW1    & 0.02 -- 0.10 \nl
Peak 1          & P1     & 0.10 -- 0.14 \nl
Trailing Wing 1 & TW1    & 0.14 -- 0.19 \nl
Interpulse 1    & IP1    & 0.19 -- 0.26 \nl
Interpulse 2    & IP2    & 0.26 -- 0.36 \nl
Bridge         	& Bridge & 0.36 -- 0.45 \nl
Leading Wing 2  & LW2    & 0.45 -- 0.50 \nl
Peak 2        	& P2     & 0.50 -- 0.57 \nl
Trailing Wing 2 & TW2    & 0.57 -- 0.64 \nl
Offpulse        & OP     & 0.64 -- 0.02 \nl
\enddata
\end{deluxetable}

The instantaneous flux of Vela shown in the center panel of
\fig{vela_hardness} exhibits a greater variation as a function of pulsar
phase than either Crab or Geminga.  The measured instantaneous flux above 100
MeV exceeds $4 \times 10^{-5}$ photons\perareasec\ for both peaks, almost 20
times the average intensity from Crab, reflective of the fact that Vela is
the brightest compact object in the \gammaray\ sky.  Although seven of the eight phase bins in the
offpulse region have a measured flux above the diffuse background,
none of the individual bins in the offpulse region measure strong detections, and one of the bins has only an upper limit.  This contrasts strongly with the Crab and Geminga, which show significant emission  over most of the pulsar phase.
A likelihood analysis of the entire offpulse phase region 0.64--0.02 finds a
point-source signal at a 5.6$\sigma$ confidence level and an instantaneous
flux of $(4.4 \pm 0.9) \times 10^{-7}$ photons\perareasec.  Thus, there is evidence that Vela has emission beyond the phase region traditionally associated with the pulsar, even though the data do not justify the claim that the emission extends throughout the full rotation.  If this level of emission
is assumed present throughout the pulse profile, it represents only $4.4\% \pm 0.9\%$
of the total emission, which does not conflict with the upper limits to
unpulsed emission established by Grenier \etal\ (1988) or Kanbach
\etal\ (1994).  Even though the flux levels measured in the offpulse regions
of the Crab and Vela pulsars are comparable, the Vela flux values have larger
uncertainties because the EGRET exposure to Vela was $\sim$ 2/3 that of the
Crab over the first three cycles of instrument operation.

The lower panel of \fig{vela_hardness} shows the distribution of hardness
ratios across the Vela pulse profile.  The weak level of offpulse emission
combined with the limited EGRET exposure to Vela prevented a useful
determination of hardness ratios over the phase range 0.85--0.05.  The phase
evolution of the Vela hardness ratio is reminiscent of the behavior of the
Crab hardness ratio (\fig{crab_hardness}).  As with the Crab pulsar, the Vela
emission becomes harder as the phase moves through the first peak and softer
as the phase increases past the second peak into the offpulse component. 

\section{Phase-Resolved Spectra}

To further investigate the implied spectral variation as a function of
rotation phase implied by the hardness ratio distributions of the three
bright \gammaray\ pulsars, differential photon spectra over the EGRET
energy range are derived for the various pulsar components
defined in Tables~\ref{crab_phasedef}, \ref{geminga_phasedef}, and
\ref{vela_phasedef}.  Each component spectrum is derived by performing
standard likelihood spatial analysis (\cite{Mattox96}) in ten independent
energy intervals using only those photons within the corresponding phase
range. As noted by Esposito et al. (1997), the EGRET in-flight calibration 
checks showed that the original calibration below 70 MeV was in error.  
The corrections necessarily introduce larger uncertainties in the 30-50 MeV 
and 50-70 MeV data points than found in the other energy bins. 

\subsection{The Crab Pulsar}

\begin{figure}[ht]
\plotone{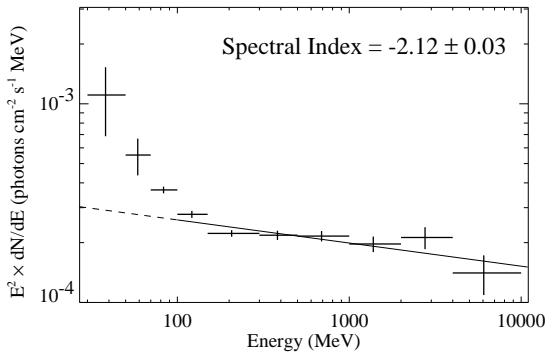}
\caption{Photon spectrum of the total emission from Crab derived using
spatial analysis.  The power law fit is to the data points between 100 MeV
and 10 GeV, with the dashed line indicating the extrapolated power law down
to 30 MeV\@.  The spectral index refers to the differential photon spectrum
$dN/dE$.
\label{crab_tot_spec}}
\end{figure}

The photon spectrum derived from likelihood spatial analysis of the total
Crab emission is shown in \fig{crab_tot_spec} for the energy range 30 MeV--10
GeV\@.  Likelihood analysis is not performed above 10 GeV because the
low photon statistics make it difficult to perform effective
maximum likelihood parameter estimation (see \cite{Mattox96}).  As indicated
by the solid line in \fig{crab_tot_spec}, the Crab differential photon
spectrum $dN/dE$ is consistent with a power law of index $\alpha = -2.12 \pm
0.03$ over the energy range 100 MeV--10 GeV\@, similar to the results of Nolan 
et al. (1993) and Ramanamurthy et al. (1995b).  However, the first three
energy bins from 30--100 MeV deviate significantly from the trend established
by the higher energy bins.  A spectral power law fit including all ten energy
bins results in a \chisq\ value of 47.5 with 8 degrees of freedom (DOF),
implying that the probability that the 10 measured values are drawn from a
photon spectrum behaving as a simple power law over the entire EGRET energy
range is $1.2 \times 10^{-7}$.  Even the restricted fit above 100 MeV
measures a \chisq\ of 10.7 for 5 DOF, corresponding to a probability of 5.9\%.

\begin{figure}
\plotone{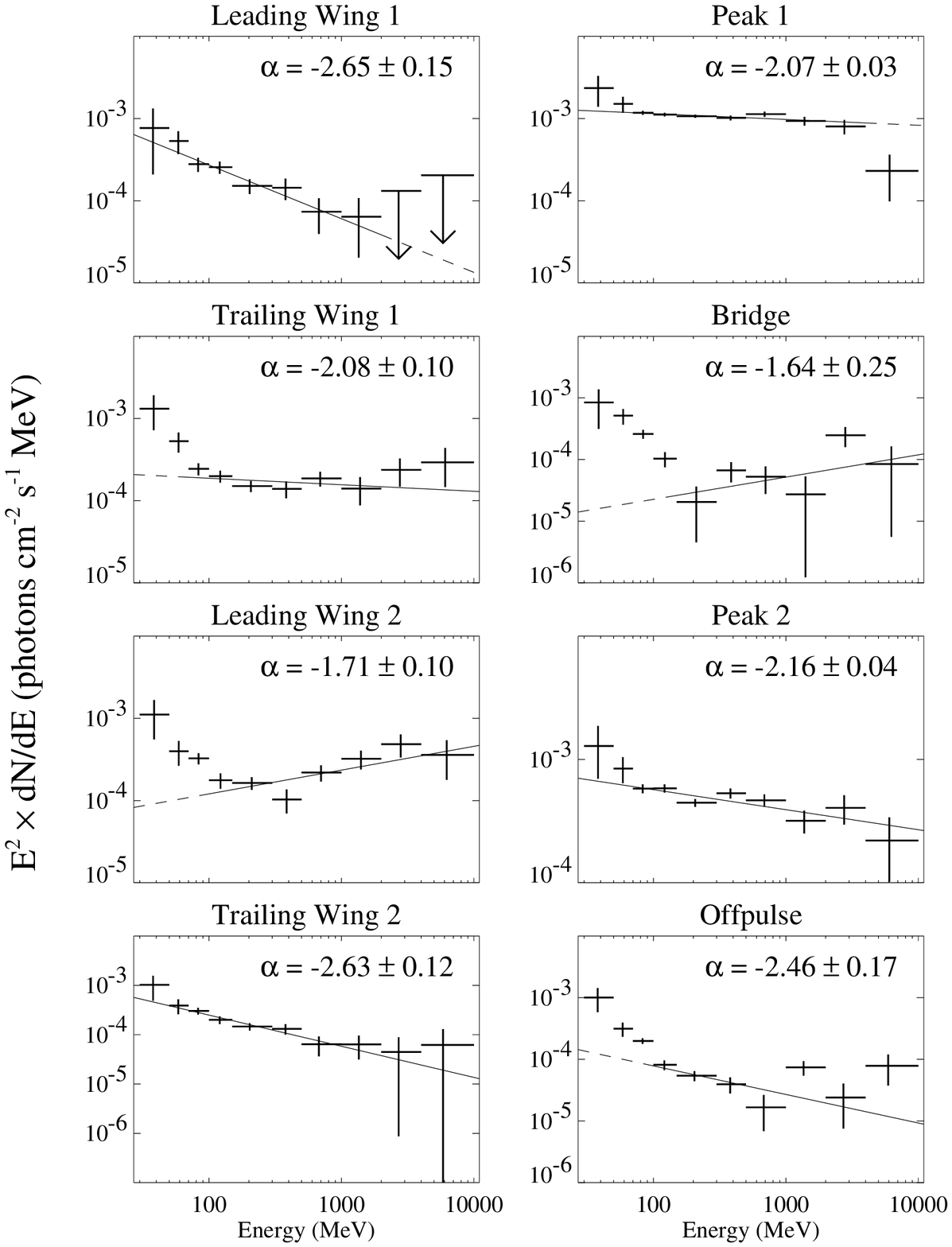}
\caption{Instantaneous photon spectra derived from likelihood spatial
analysis of the Crab pulse profile components defined in \tbl{crab_phasedef}.  
The power law fit is indicated by a solid line with the differential spectral
index $\alpha$ listed for each component.  The upper limits are $2\sigma$.
\label{crab_comp_spec} }
\end{figure}

\fig{crab_comp_spec} shows the instantaneous photon spectra derived from
likelihood analysis of the phase intervals defined in \tbl{crab_phasedef}.
The time-averaged spectra are obtained by multiplying the flux values by the
phase width of each component.  For each component, a spectral fit is only
performed over the energy interval for which the data points were reasonably
consistent with a power law distribution.  Not surprisingly, component
spectra reflect many of the features suggested by the hardness plot of
\fig{crab_hardness}.  The softest components in the Crab lightcurve are the
second pulse trailing wing (TW2), offpulse (OP), and first pulse leading wing
(LW1) components.  The pulsar emission is hardest in the bridge and leading
wing of the second pulse (LW2).  The only component to show some evidence of
a high energy turnover is the first peak (P1). Nevertheless, the absence of pulsed emission at TeV energies (e.g. \cite{Vacanti91}) indicates that the 
pulsed emission must have a cutoff between the EGRET and TeV energy ranges.  
Although based on more extensive data, the principal characteristics of 
the phase-resolved spectral components are similar to those found by 
Nolan et al. (1993) for the pulsed emission alone. 

\begin{figure}
\plotone{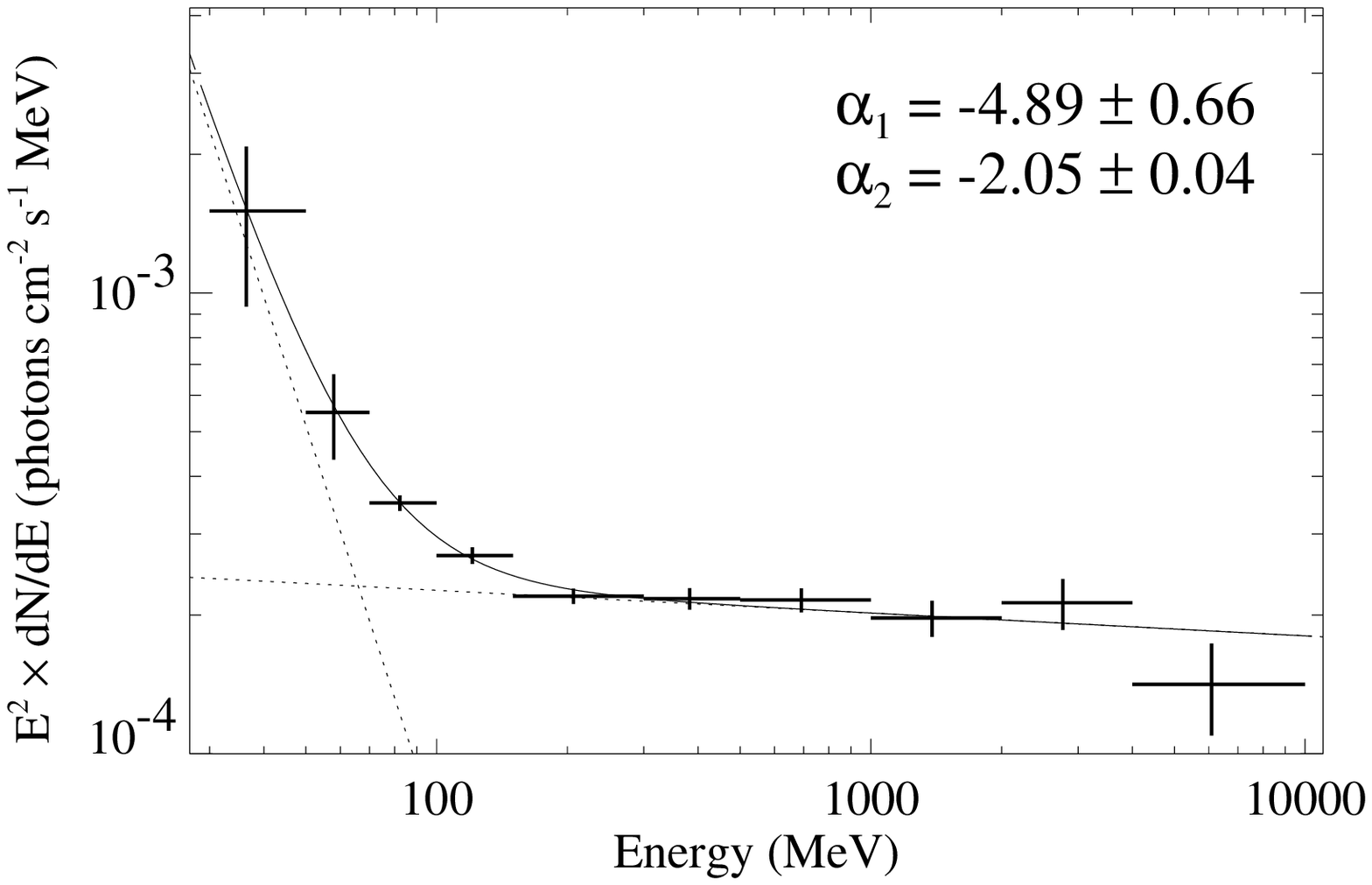}
\caption{Double power law fit to the total Crab emission photon spectrum.
The fit has a form given by \eq{double}.  The dotted lines indicate the two
independent power laws.
\label{crab_tot_dbl} }
\end{figure}

\begin{figure}
\plotone{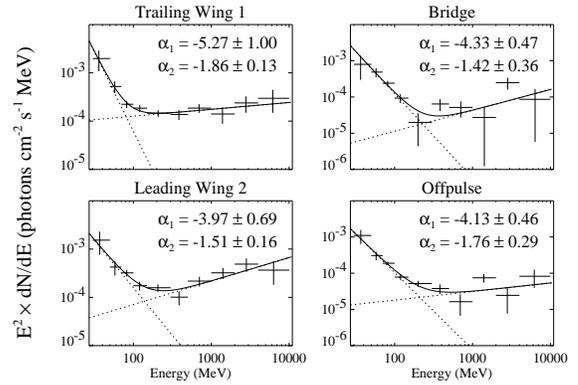}
\caption{Double power law fits to the likelihood-derived photon spectra of
the TW1, bridge,  LW2, and OP components of the Crab pulse profile.
\label{crab_comp_dbl} }
\end{figure}

Much like the total spectrum of \fig{crab_tot_spec}, the first pulse trailing
wing (TW1), bridge, LW2, and OP components all appear to have a softer
spectrum at energies less than 100 MeV which deviates from the power law
established above 100 MeV\@.  This implied spectral break is especially
apparent for the bridge and offpulse regions, the two weakest components of
the Crab pulse profile.  To investigate this behavior, the spectra are fit
with a double power law of the form
\begin{eqnarray}
\frac{dN}{dE} &= I_1 \, E^{\alpha_1} + I_2 \, E^{\alpha_2} \; \nonumber \\
&\quad {\rm ~photons~cm}^{-2}{\rm ~s}^{-1}{\rm ~MeV}^{-1}\; .
\label{double}
\end{eqnarray}
\fig{crab_tot_dbl} shows the double power law fit to the total Crab emission,
and \fig{crab_comp_dbl} shows similar fits to spectra of the TW1, bridge,
LW2, and OP components.  The various parameters of the double power law fits
are listed in \tbl{crab_dblfits}.  The \chisq\ values listed in the sixth
column indicate the measured photon spectrum for each component in
\tbl{crab_dblfits} is well-modeled by emission from two independent power
laws over the energy range 30 MeV--10 GeV\@.  The second and third columns of
\tbl{crab_dblfits} show that the soft component is consistently fit with a
spectral index of $\alpha_1 \la -4$ and a flux at 100 MeV of
$F_1({\rm100~MeV}) \sim 10^{-8}$~photons\perareasec, while the fourth and
fifth columns show that the harder high-energy emission varies for each
component.  This strongly suggests the presence of an unpulsed ultra-soft
component to the Crab total emission which dominates the pulsar radiation
from the weaker pulse profile components at energies less than $\sim$ 100
MeV\@.

\begin{deluxetable}{lcccccc}
\tablecaption{Crab Double Power Law Spectral Fits\label{crab_dblfits}}
\tablewidth{0pt}
\tablehead{
\colhead{} & \multicolumn{2}{c}{Power Law 1} & \colhead{} &
\multicolumn{2}{c}{Power Law 2} & \colhead{\chisq} \\
\cline{2-3} \cline{5-6}
\colhead{Component} & \colhead{$F_1({\rm100~MeV})$\tablenotemark{a}} &
\colhead{$\alpha_1$} & \colhead{} &
\colhead{$F_2({\rm100~MeV})$\tablenotemark{a}} & \colhead{$\alpha_2$} &
\colhead{(6 DOF)}
}
\startdata
Total  & $0.7 \pm 0.2$ & $-4.89 \pm 0.66$ & & $2.3 \pm 0.3$ & $-2.05 \pm 0.04$ & 11.04 \nl
TW1    & $0.6 \pm 0.4$ & $-5.27 \pm 1.00$ & & $1.3 \pm 0.4$ & $-1.86 \pm 0.13$ & 6.06 \nl
Bridge & $1.2 \pm 0.3$ & $-4.33 \pm 0.47$ & & $0.1 \pm 0.1$ & $-1.42 \pm 0.36$ & 10.26 \nl
LW2    & $1.6 \pm 0.4$ & $-3.97 \pm 0.69$ & & $0.7 \pm 0.4$ & $-1.51 \pm 0.16$ & 4.02 \nl
OP     & $1.0 \pm 0.2$ & $-4.13 \pm 0.46$ & & $0.2 \pm 0.2$ & $-1.76 \pm 0.29$ & 2.40 \nl
\tablenotetext{a}{Modeled instantaneous differential flux at 100 MeV in
units of $10^{-8}$~photons\perareasecmev.}
\enddata
\end{deluxetable}

\subsection{The Geminga Pulsar}
\begin{figure}
\plotone{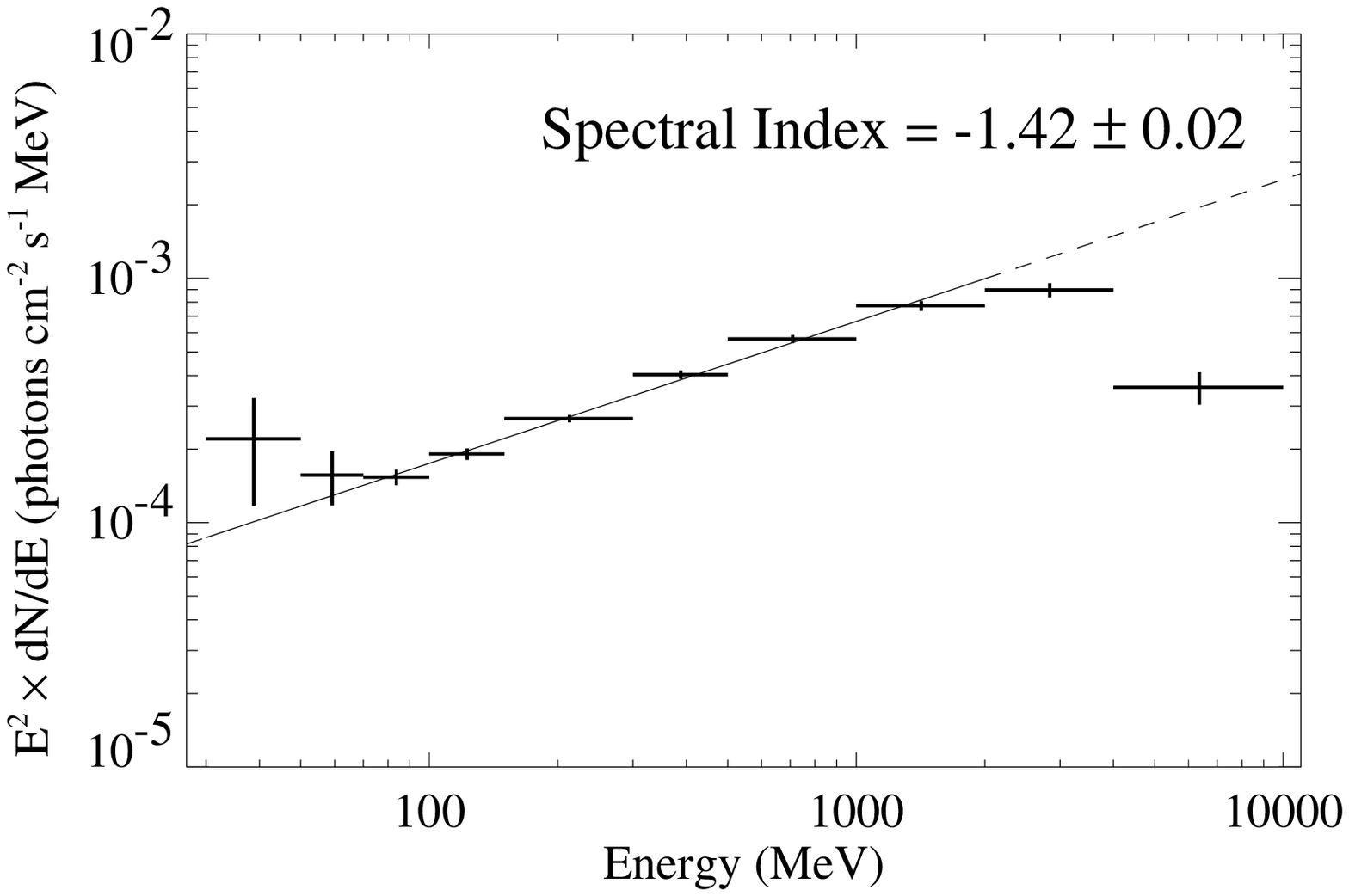}
\caption{Photon spectrum of the total emission from Geminga derived using
spatial analysis.  The power law fit is to the data points between 30 MeV and
2~GeV, with the dashed line indicating the extrapolated power law out to 10
GeV\@.
\label{geminga_tot_spec} }
\end{figure}

The total Geminga spectrum plotted in \fig{geminga_tot_spec}
is consistent with a power law from 30 MeV to 2~GeV of spectral index $-1.42 \pm 0.02$.  Besides being significantly harder and having no ultra-soft low-energy
component, the Geminga total spectrum is much different than the Crab total
spectrum in that it shows a sharp spectral turnover at high energies.  
The spectral result appears consistent with that of Mayer-Hasselwander 
et al. (1994), 
indicating that the differences in analysis technique have no strong 
influence on the results. 

\begin{figure}
\plotone{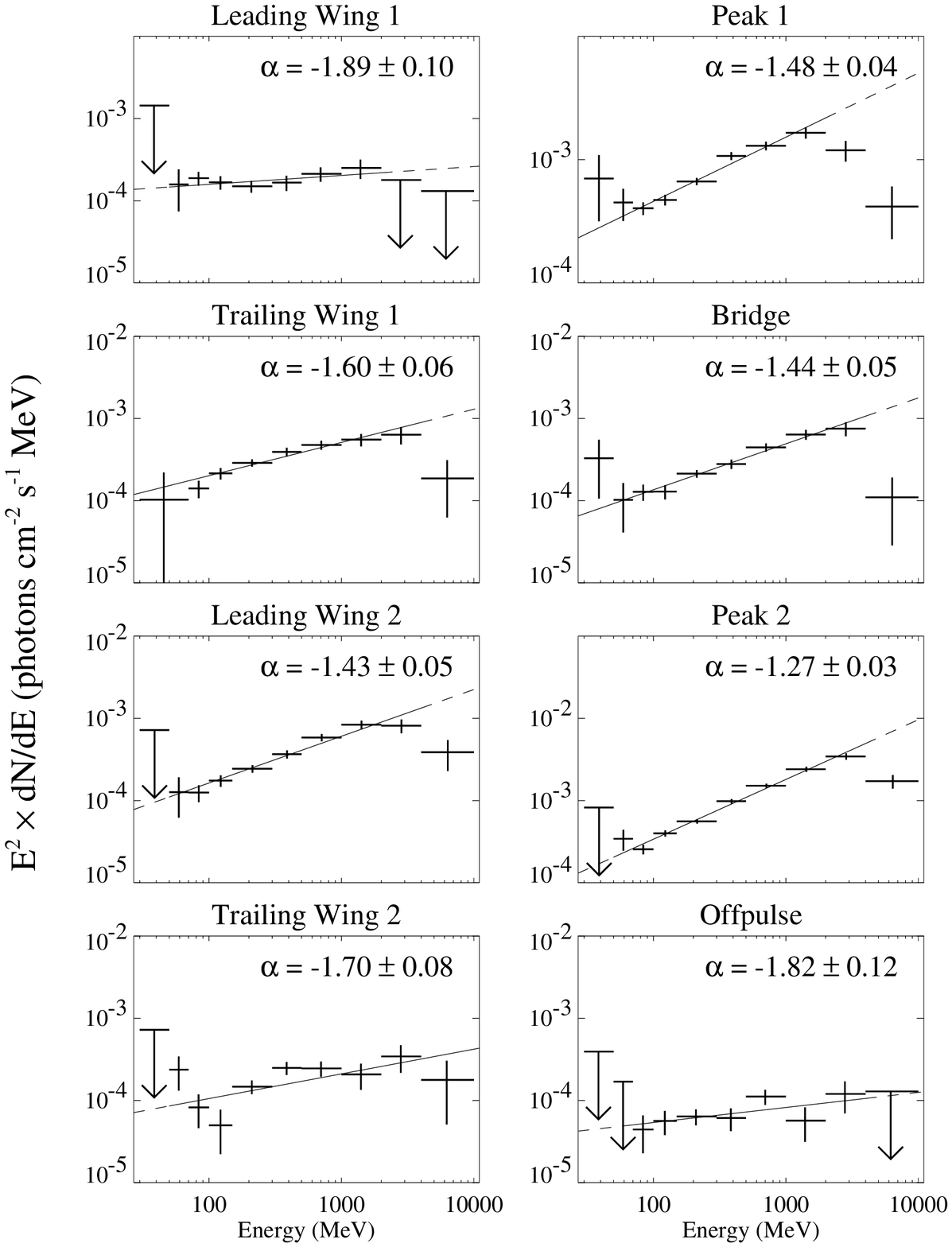}
\caption {Instantaneous photon spectra derived from likelihood spatial
analysis of the Geminga pulse profile components defined in
\tbl{geminga_phasedef}.
\label{geminga_comp_spec} }
\end{figure}

The individual component spectra of Geminga are shown in
\fig{geminga_comp_spec}.  With the exception of perhaps the TW2 and offpulse
phase intervals, all of the components show some indication of a turnover at
high energies, typically around a few GeV\@.  As is suggested by the
distribution of hardness ratios, the P2 component has a significantly flatter
spectrum than any of the other phase intervals.  Indeed, it is the hardest
spectrum yet measured by EGRET\@.  The softest components are the OP and LW1
phase intervals, which are also the only two components that are not detected
at the highest energies.  The LW1 and P1 components appear to exhibit the
earliest and strongest spectral turnovers.  Again, these results are completely compatible with those of Mayer-Hasselwander et al. (1994), although the 
definitions of phase intervals are somewhat different. 

\subsection{The Vela Pulsar}
\begin{figure}
\plotone{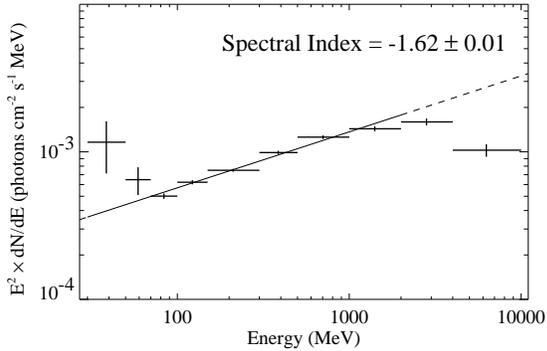}
\caption{Photon spectrum of the total emission from Vela derived using
spatial analysis.  The power law fit is to the data points between 30 MeV and
2~GeV, with the dashed line indicating the extrapolated power law out to 10
GeV\@.
\label{vela_tot_spec} }
\end{figure}

The Vela total photon spectrum is presented in \fig{vela_tot_spec}.  Over the
energy range 30--2000 MeV, the spectrum is consistent with a power low of
index $-1.62 \pm 0.01$.  Like Geminga, Vela exhibits a spectral turnover
above $\sim$ 1~GeV\@.  The shape of the spectrum is similar to that found 
by Kanbach et al. (1994), although the formal errors on the spectral index 
do not overlap.  Some of this may be attributable to time-dependent corrections 
applied to the long-term EGRET database (\cite{Esposito97}) after the 
work of Kanbach et al. (1994).  It may also reflect the fact that systematic 
uncertainties become important for a bright source like Vela, and such 
systematics are not shown in the statistical error bars.  Although it may 
appear that the Vela spectrum is
inconsistent with the fitted power law at energies below 70 MeV, the
uncertainties associated with the flux values measured for the lowest two
energy bins are large, and the deviations are not statistically significant.
 Nevertheless, the OSSE detection of unpulsed emission from the Vela synchrotron nebula
in the 0.061 to $\sim$4 MeV range (\cite{DeJager96}), 
indicates that there must be a turn-up of the total Vela spectrum near or
below the EGRET energy range.  The low-energy EGRET points are consistent 
with the cutoff of the synchrotron spectrum near 40 MeV predicted by de Jager, 
Harding, \& Strickman. 

\begin{figure}
\plotone{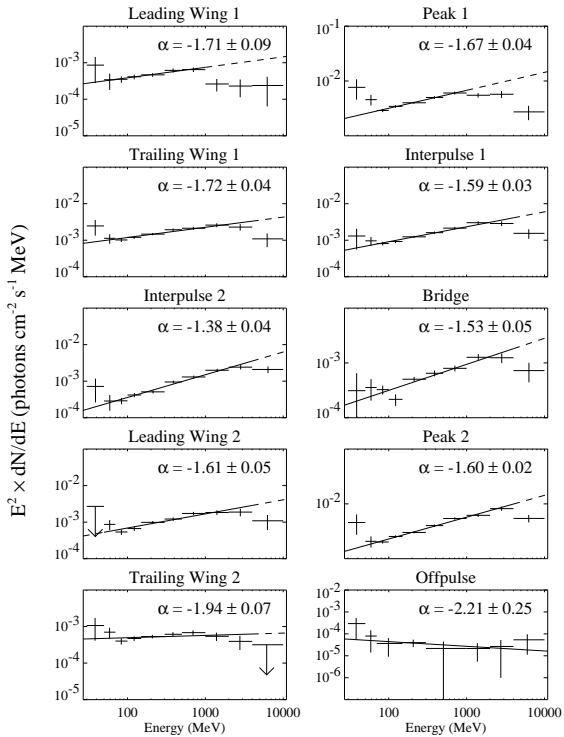}
\caption {Instantaneous photon spectra derived from likelihood spatial
analysis of the Vela pulse profile components defined in
\tbl{vela_phasedef}.
\label{vela_comp_spec} }
\end{figure}

The instantaneous differential photon spectra of the components defined in
\tbl{vela_phasedef} are presented in \fig{vela_comp_spec}.  For the OP
component, some energy bins had to be combined so that likelihood analysis
could produce a reasonable flux estimate.  Except for the OP interval, all of
the Vela pulsar components show a turnover at energies on the order of a few
GeV, with the LW1 and P1 spectra exhibiting the earliest deviations from
power law behavior.  Only the TW2 component fails to produce a signal at the
highest energies.  The IP2 phase interval is the hardest component, as was 
suggested by the hardness ratio distribution of \fig{vela_hardness}.   
Although the phase bins do not match those of Kanbach et al. (1994) exactly, 
the similarities of the two sets of phase-resolved spectra are evident. 

\section{Discussion}
\subsection{Observational Results}
A systematic, phase-resolved analysis of the total photon flux above 100 MeV reveals that
Crab, Geminga, and Vela all have significant emission in large portions
of the ``offpulse'' phase intervals.  
As opposed to studies of the Crab region,
multiwavelength observations of Geminga and Vela give no reason to believe
that their offpulse \gammaray\ emission might be due to nebular emission.  If
the Geminga and Vela pulsars are producing high-energy \gammaray\ emission
throughout much of the rotation phase, then it is reasonable to assume that at
least some of the emission in the Crab offpulse region is also generated by
the pulsar itself.  It is worth noting that for all three pulse profiles, the
lowest level of emission occurs just before the first \gammaray\
peak.  While this result is not statistically significant,
it may be an early indication that the offpulse emission is really
composed of the extended tail(s) of the major \gammaray\ peaks.

Although it is not apparent from the phase histograms shown in the top panels
of Figures~\ref{crab_hardness}, \ref{geminga_hardness}, and
\ref{vela_hardness}, phase-resolved analysis shows that the bridge emission
is different for each of the three bright pulsars.  Midway between its two
peaks, the Crab bridge emission drops to an intensity level which is
consistent with the weak emission in the offpulse region.  The Vela bridge
emission also varies substantially, but remains at a consistently high flux
level and has a minimum very near the second peak.  The emission from Geminga
is both strong and relatively steady throughout the bridge region.

Despite the reduced photon statistics incurred by phase-resolved analysis, it
is possible to acquire spectral information in most of the twenty phase bins
by employing a hardness ratio such as that defined in \eq{hardness}.  The
Crab and Vela spectral variation are qualitatively similar in that the
hardness ratios increase smoothly through the first peak, have maximum values
near the bridge region, and drop off after the second pulse.  
The Vela hardness ratios are typically greater than those of the Crab
pulsar. Also, Vela becomes dramatically harder after the first pulse, an
effect not seen in the Crab hardness ratios.  In contrast to Crab and Vela,
the hardness ratio for Geminga is strongly correlated with the features in
its pulse profile.  There are two broad plateaus corresponding to the first
peak and bridge region, with an even larger hardness ratio at the location of
the second peak.  For all three pulsars, the softest ratios are in the
offpulse region.

Spectra derived from likelihood analysis of the individual pulse profile
components confirm the behavior suggested by the distribution of hardness
ratios.  Of the three bright \gammaray\ pulsars, the softest emission occurs
in the leading wing of the first Crab pulse which has a measured spectral
index of $-2.65 \pm 0.15$, with an almost identical spectral index measured
for the trailing wing of the second pulse.  The hardest pulsar component is
the second peak from Geminga, which has a measured spectral index of $-1.27
\pm 0.03$.  The latter part of the Vela interpulse component is almost as
hard with a measured spectral index of $-1.38 \pm 0.04$.

For all three pulsars, the component spectra share the same basic features as
the total spectra.  That is to say, the high-energy spectral turnovers which
are apparent in the Vela and Geminga total spectra also are also evident in
the various component spectra.  Likewise, the Crab profile components show
the same soft emission at low energies that is apparent in the total Crab
spectra, and only the leading peak shows some evidence of a high-energy
turnover in its photon spectrum.  Interestingly, for both the Vela and
Geminga pulsars, the leading peak spectrum similarly turns over at a lower
energy than the other pulsar components.

The four weakest Crab pulsar components are dominated by ultra-soft emission
at energies below $\sim$100 MeV\@.  Double power law fits to the photon
spectra from these four components show that the soft emission appears at
roughly the same level in each component, with an average spectral index of
$-4.3 \pm 0.3$ and a flux of $(1.1 \pm 0.1) \times 10^{-8}$
photons\perareasecmev\ at 100 MeV\@.  This ultra-soft emission is also
apparent in a double power law fit to the total Crab emission, clearly
indicating that it is an unpulsed component to the Crab \gammaray\ emission.
It is not unreasonable to expect ultra-soft nebular emission out to energies
on the order of $\sim 100$~MeV\@.  The emission at X-ray and soft
\gammaray\ energies is consistent with synchrotron radiation produced by
relativistic electrons streaming from the pulsar (\cite{Kennel84}).  Since
the Crab pulsar is unlikely to produce electrons with energies greater than
$\sim 10^{15}$--$10^{16}$ eV, the synchrotron spectrum will begin to rollover
at energies above a few MeV\@.  The soft emission observed from 30--100 MeV
is then seen as the tail end of the synchrotron emission from the nebula.  
Detailed modeling of the Crab emission by de Jager et al. (1996) confirms 
this qualitative picture. 

The detection of unpulsed emission from the Crab complex at TeV energies
(\cite{Weekes89}) implies that there is emission from the nebula beyond the
synchrotron radiation.  The TeV emission is explainable in terms of
inverse-Compton scattering (\cite{deJager96}), which has also been proposed
as a possible explanation of the harder-spectrum offpulse emission above 1
GeV\@.  However, based on the level of the TeV emission, the expected
inverse-Compton contribution at energies on the order of $\sim$1 GeV is below
the EGRET flux measurements (\cite{Aharonian95}).  Considering the non-zero
offpulse emission seen from the Geminga and Vela pulsars, it is very likely 
that the bulk of  the hard power-law component of the Crab
offpulse emission is generated by the pulsar itself.

\subsection{Theoretical Implications}
Original polar cap models of pulsar \gammaray\ emission (\eg\
\cite{Harding81}) suggested that double-peaked pulse profiles were produces
by pulsars whose rotation axis was nearly orthogonal to its dipole axis.
However, this could not explain peak separations much different than 0.5 in
rotation phase or strong bridge emission between peaks.  Since at least five
of the six high-energy \gammaray\ pulsars have more than one peak
(\cite{Ramanamurthy95a}; \cite{Fierro95}; \cite{Thompson96}), it is unlikely
that these are all orthogonal rotators whose beams cross the line of sight to
the Earth.  More recent incarnations of the polar cap models (\eg\
\cite{Dermer94}; \cite{Sturner94}; \cite{Daugherty94}) have made use of the
fact that a pulsar with closely aligned rotation and magnetic axes pointed in
the direction of the Earth will produce a hollow cone of emission from a
single polar cap and result in two observed peaks, corresponding to when the
line of sight from the Earth enters and exits the hollow cone.  Yet
phase-resolved analysis shows that all three bright \gammaray\ pulsars have
emission across much of the rotation phase, which for the single
pole outer gap model demands that the rotation and dipole axes have an aspect
angle no greater than the radius of the polar cap cone of emission.  For the
narrow beaming angles associated with the standard polar cap model (\eg\ 
\cite{Sturrock71}), this suggests that there is an very large population of
young nearby neutron stars with closely aligned rotation and dipole axes.

Daugherty \& Harding (1996) refined the single polar cap model by allowing the
accelerating regions above the polar caps to extend out to heights on the
order of a neutron star radius, significantly increasing the solid angle into
which the cone of \gammaray\ emission is beamed.  Not only does this make it
more likely that a pulsar beam will cross the line of sight from the Earth,
but it relaxes the requirement that the pulsar rotation and dipole axes be
nearly aligned.  Under these assumptions, Daugherty \& Harding (1996) are
able to closely reproduce the Vela pulse profile observed by EGRET, matching
the narrow peak duty cycles, observed peak separation, and enhanced bridge
emission.  They also predict there will be a finite level of emission outside
of the two peaks due to residual high-altitude cascades.  A natural
consequence of this scenario is that the bridge emission will be harder than
the peak emission since it comes from the interior of the polar cap where the
curvature radiation is less likely to cross magnetic field lines and be
reprocessed into softer cascade photons.  The emission outside of the peaks
is due to high-altitude curvature radiation from electrons which have lost
much of their peak energy, so this emission is expected to be the softest.
It is also expected that the photon spectra from the peaks will turn over at
lower energies than the spectra from the bridge regions since the peak
emission will be more heavily attenuated by the magnetic field.  It has
already been noted that the leading peaks of the Crab, Vela, and Geminga
pulsars have spectra which appear to turn over at lower energies than the
other pulsar components.  While the measured hardness ratio distributions of
the Crab and Vela pulsars qualitatively agree with the behavior expected for
this polar cap scenario, the phase-resolved spectral behavior of the Geminga
pulsar presents some problems.  By far the hardest emission from Geminga
comes from the second peak.  Moreover, the emission between the peaks shows a
sharper high-energy spectral turnover than the second peak.  Both of these
conditions indicate that the emission producing the second Geminga peak is
not being significantly reprocessed to softer photons by the magnetic field.

The emission pattern of the magnetic Compton-induced pair cascade model
(\cite{Dermer94}; \cite{Sturner94}) is similar to that of the curvature
radiation-induced pair cascade model of Daugherty \& Harding (1996), except
the high-energy \gammaray\ emission is produced close to the surface of the
star, resulting in a much narrower \gammaray\ beam.  In calculating the
\gammaray\ spectral index as a function of pulsar phase, Sturner, Dermer, \&
Michel (1995) find that the hardest emission should come from the bridge
region and the softest emission from the two peaks.  This agrees somewhat
with the observational results for the Crab and Vela pulsars---although the
observed hardness ratios are not directly correlated with the peaks as
suggested by Sturner \etal\ (1995)---but is in conflict with the extremely
hard photon spectrum measured for the second \gammaray\ peak of Geminga.
Moreover, the pulse profiles modeled by Sturner \etal\ (1995) do not reflect
the narrow duty cycles of the peaks,  strong bridge emission, nor the
significant offpulse emission observed from the three brightest \gammaray\ 
pulsars.

Chiang \& Romani (1992) found that outer gap emission from a single pole
produces a broad, irregularly-shaped beam of emission which is particularly
dense near the edges, so that two \gammaray\ peaks are observed when the line
of sight from the Earth crosses these enhanced regions of the
\gammaray\ beam, while the inner region of the beam provides a significant
amount of emission between the peaks.  A wide range of peak phase separations
can be accommodated with a proper choice of the observer co-latitude.
Double-peaked pulse profiles of varying phase separation and strong bridge
emission occur naturally in this model.  The non-zero offpulse emission can
be seen as resulting from residual pair cascades high in the outer
magnetosphere Due to the complex geometry and interaction of emission
processes in the outer magnetosphere, no detailed spectra or luminosities
have been modeled for single pole outer gap model.

A clue to the nature of pulsar \gammaray\ emission may come from the fact
that for each of the three bright \gammaray\ pulsars, the individual
component spectra have comparable high-energy turnovers, if any.  Magnetic
field attenuation would be a likely cause of the spectral turnovers if the
the various pulsar components are being generated close to the surface of the
neutron star, where the local magnetic field is strong.  However, if the
turnovers are due to attenuation by the magnetic field, the Crab pulsar with
an inferred surface magnetic field of $3.7 \times 10^{12}$~G should show a
sharper spectral break than a weaker field pulsar like Geminga, which has an
inferred surface magnetic field of $1.6 \times 10^{12}$~G\@.  It is also possible that the spectral turnovers are the result of a cutoff in the source
distribution of charged particles at high energies.  Assuming the maximum
energy attainable by a charged particle is directly related to the maximum
potential drop of the pulsar, which goes as $\Delta\Phi \propto B P^{-2}$,
where $B$ is the surface magnetic field and $P$ is the rotation period of the
pulsar, one would expect the Crab pulsar to have the highest charged particle
distribution turnover energy among the three bright \gammaray\ pulsars, while
Geminga should have the lowest.  This could be why the Crab photon spectrum 
shows no strong spectral turnover below 10 GeV, while Geminga has a sharp 
turnover. On the other hand, PSR B1951+32, with a  $\Delta\Phi$ lower than 
Vela's, shows no high-energy cutoff (\cite{Ramanamurthy95a}).

A spectral turnover caused by a dropoff in the distribution of primary
charged particles at high energies will be reflected in both the total
emission photon spectrum and the individual component spectra.  However, it
is not clear why the first peak of the the \gammaray\ pulsars seems to have a
slightly stronger cutoff than do the other pulsar components.  Perhaps the
first peak originates closer to the surface of the pulsar where the local
magnetic field is stronger and there is a higher degree of magnetic field
attenuation, but then it might be expected that the first peak should be
softer than the other components due to the significant reprocessing of
high-energy photons to lower energies.  While Vela and Geminga have somewhat
softer leading peaks, the Crab pulsar does not.

Even though there is some question as to the exact location in the
magnetosphere where each pulse profile component is generated, the
significant variation in hardness ratio as a function of rotation phase
indicates the line of sight to the pulsar is sweeping across field lines with
a broad range in curvature radii.  Most likely, the softest components come
from the regions with the most curved field lines.  The two peaks of the Crab
and Vela pulsars are softer than the bridge regions, which suggests that the
peaks might be at least partially due to the increase in curvature radiation
one would expect along the most curved field lines.  Yet, the second peak of
Geminga, which is actually the stronger peak, has the hardest measured
spectral index of any object or pulsar component observed by EGRET\@.  Either
the primary charged particle distribution responsible for this peak has a
very enhanced, flat number spectrum, or, more plausibly, the second Geminga
peak results from inverse-Compton scattering, which can reasonably produce an
intense, hard photon spectrum.  The second peak of Geminga appears to be
direct evidence that processes besides curvature radiation and synchrotron
radiation are playing an important role in the pulsar magnetosphere.

\acknowledgments
The authors would like to thank Roger Romani for his helpful discussions
regarding the soft emission in the Crab total spectrum.  We thank the referee, Hans Mayer-Hasselwander, for his helpful comments. 

The EGRET team gratefully acknowledges support from the following:
NASA Grant NAG5-1605 (SU),
NASA Cooperative Agreement NCC 5-95 (HSC), 
NASA Contract NAS5-96051 (NGC),
Bundesministerium f\"ur Bildung, Wissenschaft, Forschung und Technologie 
grant 50 QV 9095 (MPE), and
Deutsche Forschungsgemeinschaft (Sonderforschungsbereich 328).

\end{document}